\documentclass[times,twocolumn]{aastex631}
\usepackage{amssymb}
\usepackage{amsmath}
\usepackage{gensymb}
\usepackage{graphicx}
\usepackage{xcolor}
\usepackage{natbib}

\newcommand{\beq}{\begin{equation}}
\newcommand{\eeq}{\end{equation}}

\newcommand{\Ms}{{\rm{M}_*}}
\newcommand{\Msun}{\rm{M}_\odot}
\newcommand{\kmps}{km~s$^{-1}$}
\newcommand{\MHI}{\rm{M_{H{\textsc i}}}}

\newcommand{\htwo}{${\rm H_2}$}

\newcommand{\hi}{H{\sc i}}

\newcommand{\hii}{H{\sc i} 21\,cm}
\newcommand{\MHtwo}{\rm{M_{H_2}}}

\newcommand{\Mat}{{\rm{M_{Atom}}}}
\newcommand{\Mmol}{{\rm{M_{Mol}}}}
\newcommand{\Mati}{{\rm{M_{Atom,i}}}}
\newcommand{\Mmoli}{{\rm{M_{Mol,i}}}}
\newcommand{\Matf}{{\rm{M_{Atom,f}}}}
\newcommand{\Mmolf}{{\rm{M_{Mol,f}}}}
\newcommand{\Msi}{{\textrm{M}_{*,i}}}
\newcommand{\Msf}{{\textrm{M}_{*,f}}}

\newcommand{\Racc}{\rm{R_{Acc}}}
\newcommand{\Rmol}{\rm{R_{Mol}}}
\newcommand{\Rsfr}{\rm{R_{SF}}}

\graphicspath{{./}{figures/}}

\shorttitle{The Gas Accretion Rate at $z\approx1$}
\shortauthors{Chowdhury, Kanekar and Chengalur}

\begin{document}

\title{The Gas Accretion Rate of Galaxies over $z\approx0-1.3$}	

	\correspondingauthor{Nissim Kanekar}
	\email{nkanekar@ncra.tifr.res.in}
	
	\author{Aditya Chowdhury}
	\affil{National Centre for Radio Astrophysics, Tata Institute of Fundamental Research, Pune, India.}
	
	\author{Nissim Kanekar}
	\affil{National Centre for Radio Astrophysics, Tata Institute of Fundamental Research, Pune, India.}
	
	\author{Jayaram N. Chengalur}
	\affil{National Centre for Radio Astrophysics, Tata Institute of Fundamental Research, Pune, India.}

	
	
\begin{abstract}
We present here estimates of the average rates of accretion of neutral gas onto main-sequence galaxies and the conversion of atomic gas to molecular gas in these galaxies at two key epochs in galaxy evolution: (i)~$z\approx1.3-1.0$, towards the end of the epoch of peak star-formation activity in the Universe, and (ii)~$z\approx1-0$, when the star-formation activity declines by an order of magnitude. We determine the net gas accretion rate $\rm{R_{Acc}}$ and the molecular gas formation rate $\rm{R_{Mol}}$ by combining the relations between the stellar mass and the atomic gas mass, the molecular gas mass, and the star-formation rate (SFR) at three epochs, $z=1.3$, $z=1.0$, and $z=0$, with the assumption that galaxies evolve continuously on the star-forming main-sequence. We find that, for all galaxies, $\rm{R_{Acc}}$ is far lower than the average SFR $\rm{R_{SFR}}$ at $z\approx1.3-1.0$; however, $\rm{R_{Mol}}$ is similar to $\rm{R_{SFR}}$ during this interval. Conversely, both $\rm{R_{Mol}}$ and $\rm{R_{Acc}}$ are significantly lower than $\rm{R_{SFR}}$ over the later interval, $z\approx1-0$. We find that massive main-sequence galaxies had already acquired most of their present-day baryonic mass by $z\approx1.3$. At $z\approx1.3-1.0$, the rapid conversion of the existing atomic gas to molecular gas was sufficient to maintain a high average SFR, despite the low net gas accretion rate. However, at later times, the combination of the lower net gas accretion rate and the lower molecular gas formation rate leads to a decline in the fuel available for star-formation, and results in the observed decrease in the SFR density of the Universe over the last 8~Gyr.

	\end{abstract}
	
	
\section{Introduction}

The evolution of galaxies is driven by the baryon cycle in which the baryonic constituents of galaxies and their circumgalactic mediums (CGMs) interact with each other and convert from one form to another \citep[e.g.][]{Peroux20}. The key processes in the baryon cycle are (i)~the accretion of mostly ionized gas from the CGM onto the ``disks'' of galaxies, forming neutral atomic hydrogen (\hi) in the disks, (ii)~the cooling of \hi\ and its conversion to molecular hydrogen (\htwo), (iii)~the gravitational collapse and fragmentation of molecular clouds to form stars, and (iv)~the expulsion of gas from the interstellar mediums (ISMs) of galaxies in outflows driven by stars or active galactic nuclei (AGNs), with some fraction of this gas later returning to the galaxy disk.

Measurements of the rates at which the above processes occur in galaxies, how the rates compare to each other, and how they evolve with redshift are critical to understanding galaxy evolution. Unfortunately, only the redshift evolution of the stellar properties of galaxies and their star-formation rates (SFR) are well determined today. For example, we have known for over two decades that the SFR density of the Universe peaks in the redshift range $z\approx1-3$ and then declines by an order of magnitude from $z\approx1$ to $z\approx0$ \citep[e.g.][]{Madau14}. Further,  $\approx90\%$ of the star-formation activity of the Universe out to $z\approx2$ occurs on the ``star-forming main sequence'' \citep[e.g.][]{Rodighiero11} --- a tight relationship between the SFR and the stellar mass ($\Ms$) of galaxies \citep[e.g.][]{Noeske07,Whitaker14}. At a fixed stellar mass, the SFR of galaxies on the main sequence declines by a factor of $\approx10$ from $z\approx1$ to the present time \citep[e.g.][]{Whitaker14,Popesso22}.

Unlike the SFR of a galaxy, it is very challenging to observationally determine the gas accretion rate and the rate at which \hi\ is converted to \htwo. {While there have been suggestions in the literature that a low gas accretion rate at late times might account for various observational results \citep[e.g.][]{Bouche10,Moller13,Walter20,Chowdhury20}, the lack of actual measurements of the accretion rate} has been a key limitation in our understanding of the processes that drive the redshift evolution of the star-formation activity of the Universe. For example, \citet{Walter20} used the redshift evolution of the cosmological SFR density, the cosmological \hi\ mass density, and the cosmological \htwo\ mass density to estimate the rate of global flow of gas, and global conversion of \hi\ to \htwo. However, such global rates, averaged over cosmological volumes, are difficult to interpret and do not provide any information on the differences between different galaxy populations, such as the dependence of the accretion rate on galaxy stellar mass, environment, morphology, etc. {Conversely, \citet{Bouche10} used a toy ``reservoir'' model, in which gas accretion is quenched above a halo mass of $\approx 10^{11} \, \Msun$, to argue that the decline in SFR density from $z \approx 2$ may be driven by the decline in the gas accretion rate. This model was found to yield the observed main-sequence and Tully-Fisher scaling relations at different redshifts \citep[see also][for the mass-metallicity relation]{Moller13}. However, its predicted gas fractions ($\approx 30-45$\% at $z \approx 1.2$) are lower than the values recently measured ($\approx 80$\% at $z \approx 1.3$) by \citet{Chowdhury22c}.}

The net gas accretion rate (i.e. the difference between the gas inflow and outflow rates) and the \htwo\ formation rate  can be determined if one knows the dependences of the \hi\ and \htwo\ masses of galaxies on their stellar masses \citep[i.e. the $\MHI - \Ms$ and $\Mmol - \Ms$ scaling relations; ][]{Scoville17,Bera23b}, along with the standard assumption that star-forming galaxies evolve along the main sequence \citep[e.g.][]{Renzini09,Peng10,Leitner12,Speagle14,Scoville17}. {For the molecular component, various observational tracers (primarily CO rotational lines and dust continuum emission)  have been used to determine the $\Mmol - \Ms$ relation of galaxies out to $z\approx 5$ \citep[e.g.][]{Tacconi20}.}
However, the weakness of the \hii\ line, the only tracer of the \hi\ mass of galaxies, has meant that, until very recently, estimates of the $\MHI - \Ms$ relation were limited to the local Universe \citep[e.g.][]{Catinella18,Parkash18}. 

The \hii\ stacking approach, based on combining the \hii\ emission signals from a large number of galaxies with accurately known positions and redshifts, allows one to overcome the intrinsic weakness of the \hii\ line and measure the average \hi\ properties of galaxy populations at cosmological distances \citep{Zwaan00,Chengalur01}. This approach has been recently used to measure the \hi\ properties of star-forming galaxies out to $z\approx1$ \citep[e.g.][]{Bera19,Bera22,Bera23,Chowdhury20,Chowdhury21,Chowdhury22b}. 
Recently, \citet{Chowdhury22d} applied the \hii\ stacking approach to data from the Giant Metrewave Radio Telescope (GMRT) Cold-\hi\ AT $z\approx1$ (CAT$z1$) survey, a 510-hr \hii\ emission survey of galaxies at $z=0.74-1.45$ \citep{Chowdhury22b}, to obtain the first measurement of the $\MHI - \Ms$ scaling relation at $z\approx1$.

In this Letter, we combine measurements of the $\MHI - \Ms$ scaling relation at $z \approx 1$ from the GMRT-CAT$z1$ survey with estimates of the star-forming main-sequence relation and the $\Mmol - \Ms$ scaling relation from the literature, and the assumption of the continuity of galaxy evolution along the main sequence, to determine the average \htwo\ formation rate and the average net gas accretion rate in star-forming galaxies over the redshift intervals $z\approx 1.3-1.0$ and $z\approx1.0-0$. 

Throughout this work, we assume a Chabrier initial mass function (IMF) for estimates of the stellar masses and SFRs. Further, we use a flat Lambda-cold dark matter cosmology, with $\Omega_m=0.3$, $\Omega_\Lambda = 0.7$, and $H_0 = 70$~\kmps~Mpc$^{-1}$.

\section{Determination of the Gas Accretion Rate and the \htwo\ Formation Rate}
\subsection{Formalism}
\label{sec:formalism}
The formalism used in this work to determine the net gas accretion rate and the \htwo\ formation rate of star-forming galaxies was introduced by \citet{Scoville17}, and recently refined by \citet{Bera23b} to include the $\MHI - \Ms$ relation. In this approach, the build-up of the stellar mass of a galaxy, with an initial stellar mass ($\Msi$) at the initial epoch ($t_i$), is tracked to determine the final stellar mass ($\Msf$) at the final epoch ($t_f$). This is done by assuming that the galaxy remains on the star-forming main-sequence over the entire period, such that 
\begin{equation}
\label{eqn:Msbuildup}
\Msf=\Msi+\int_{t_i}^{t_f} (1-f_{return})\  \mathrm{SFR_{MS}}(\Ms,t)\ \mathrm{dt} \;
\end{equation} 
where $\mathrm{SFR_{MS}}(\Ms,t)$  is the main-sequence relation at the time $t$, and $f_{return}$ is the fraction of the stellar mass returned to the ISM via stellar winds or supernovae \citep[$f_{return}= 0.41$ for a Chabrier IMF;][]{Leitner11,Madau14}. We define here the average star-formation rate $\Rsfr$ between the epochs $t_i$ and $t_f$, such that
\begin{equation}
\label{eqn:stellar}
    \Msf=\Msi+(1-f_{return})\  {\Rsfr} \Delta t
\end{equation}
where $\Delta t=t_f-t_i$ is the time interval between the two epochs of interest. Measurements of the star-forming main-sequence \citep[e.g.][]{Whitaker14,Popesso22} can then be used to track the evolution of the stellar mass of a main-sequence galaxy using Equation~\ref{eqn:Msbuildup}, and this can be combined with Equation~\ref{eqn:stellar} to determine $\Rsfr$ using the following relation.

\begin{equation}
\label{eqn:rsfr}
{\Rsfr} =\frac{(\Msf-\Msi)}{(1-f_{return})\Delta t} \; 
\end{equation}

Next, the final molecular gas reservoir\footnote{Throughout this Letter, $\Mmol$ and $\Mat$ are used to refer to gas masses that include the mass contribution of Helium}, $\Mmolf$, of the galaxy at time $t_f$  is related to the initial molecular gas reservoir, $\Mmoli$, at time $t_i$  via 
\begin{equation}
    \label{eqn:molgas}
    \Mmolf=\Mmoli - \Rsfr \Delta t + \Rmol \Delta t
\end{equation}
where $\Rmol$ is the average molecular gas formation rate between the two epochs of interest, i.e. the difference between the rate at which \hi\ is converted to \htwo\ and that at which \htwo\ is dissociated to \hi. Given that the initial and final stellar masses are known, one can use measurements of the molecular gas mass as a function of stellar mass at both epochs to determine $\Mmolf$ and $\Mmoli$ which can then be combined with Equation~\ref{eqn:molgas} to obtain $\Rmol$ using the following relation.

\begin{equation}
\label{eqn:rhtwo}
\Rmol=\Rsfr+(\Mmolf-\Mmoli)/\Delta t
\end{equation}

Finally, the neutral atomic gas mass, $\Matf$, of the galaxy at time $t_f$ is related to the initial neutral atomic gas mass, $\Mati$, at time $t_i$ via the relation
\begin{equation}
    \label{eqn:atgas}
    \Matf=\Mati - \Rmol \Delta t + \Racc \Delta t
\end{equation}
where $\Racc$ is the \emph{net} accretion rate of neutral atomic gas onto the disk of the galaxy, i.e. the difference between the neutral atomic gas inflow and outflow rates. Again, measurements of $\MHI$ as a function of stellar mass at the two epochs  can be used to determine $\Matf$ and $\Mati$, which can be combined with Equation~\ref{eqn:atgas} to obtain $\Racc$ using the following relation:

\begin{equation}
\label{eqn:racc}
\Racc=\Rmol +(\Matf-\Mati)/\Delta t
\end{equation}

We note that in the above formalism, the stellar mass returned to the ISM is assumed to be mostly in the ionised state and is hence not explicitly included in Equations~\ref{eqn:molgas}-\ref{eqn:racc}. It is entirely possible that some fraction of this gas cools to form \hi\ or \htwo\ over the time interval between $t_i$ and $t_f$. However, we emphasize that even assuming that \emph{all} of the stellar mass that is returned to the ISM cools to form \hi\ would not significantly alter the key conclusions of this work.

We use the above formalism to estimate $\Rsfr$, $\Rmol$, and $\Racc$ over two redshift intervals: (a) $z\approx1.3$ to $z \approx 1.0$, covering the end of the epoch of peak cosmic star-formation activity, and (b) $z\approx1.0$ to $z\approx0$, covering the decline of star-formation activity over the last $\approx 8$~Gyr \citep[e.g.][]{Madau14}. The choice of these redshift intervals is driven by our recent measurements of the average \hi\ content of stellar mass-matched samples of star-forming galaxies at $z \approx 1.0$ and $z \approx 1.3$ \citep{Chowdhury22a}, and the measurement of the $\MHI - \Ms$ scaling relation over the redshift range $0.74 \leq z \leq 1.45$ \citep{Chowdhury22d}. In the following subsections, we briefly describe the relations that were used to estimate SFR, $\Mmol$, and $\Mat$ for star-forming galaxies at $z\approx1.3$, $z\approx1.0$, and $z\approx 0$, specifically (i) the star-forming main-sequence relation \citep[e.g.][]{Whitaker14,Popesso22}, (ii)~the $\Mmol-\Ms$ relation \citep{Tacconi20}, and (iii)~the $\MHI-\Ms$ relation \citep{Catinella18,Chowdhury22d}.

\subsection{The Star-Forming Main-Sequence}
\label{sec:sfr}

We use the star-forming main-sequence relation of \citet{Popesso22} to estimate the build-up of the stellar mass of galaxies via Equation~\ref{eqn:Msbuildup}. These authors converted a large number of literature measurements of the main sequence at $0 < z < 6$ to a common calibration, to find that the data are consistent with a relation that is linear at low stellar masses and flattens above a characteristic stellar mass \citep[see also][]{Whitaker14}. We will use the following main-sequence relation, obtained via a second-order polynomial fit to the dependence of $\log[{\rm SFR}]$ on $\log [\Ms]$  \citep{Popesso22}:

\begin{equation}
\label{eqn:popesso}
{\log}[{\rm SFR} (\Ms,t)] = (a_1 t + b_1 ) {\log} [\Ms] + b_2 {\log}^2[\Ms] + (b_0 + a_0 t)
\end{equation}
where $t$ is the age of the Universe at the epoch of interest and the fit parameters are $a_0=0.20\pm0.02$, $a_1=-0.034\pm0.002$, $b_0=-26.134\pm0.015$, $b_1=4.722\pm0.012$, and $b_2=-0.1925\pm0.0011$ \citep{Popesso22}. The main-sequence relations at $z\approx0$, $z\approx1.0$, and $z\approx1.3$, obtained using Equation~\ref{eqn:popesso}, are shown as the solid blue curves in Figure~\ref{fig:fig1}[A].

In passing, we emphasize that the main conclusions of this work are unchanged on using the second main-sequence relation of \citet{Popesso22}, which asymptotically approaches a maximum SFR value with increasing stellar mass, or the main-sequence relation of \citet{Whitaker14}, which is commonly used in the literature.

\subsection{The $\Mmol-\Ms$ Relation}
\label{sec:h2}
For the $\Mmol - \Ms$ relation, we follow \citet{Tacconi20}, who compiled literature estimates of the molecular gas mass of 2052 galaxies with $\Ms\approx10^9-10^{11.5}~\Msun$ at $0\lesssim z\lesssim5.2$. {The molecular gas masses were estimated using different tracers such as the CO rotational lines \citep[e.g.][]{ Daddi10,Tacconi13,Saintonge17}, the far-infrared dust continuum \citep[e.g.][]{Santini14}, and the 1~mm dust continuum \citep[e.g.][]{Scoville16}}. \citet{Tacconi20} corrected the measurements obtained from the different tracers for any zero-point offsets, to put them on the same scale, and obtained the following relation for the dependence of the molecular gas mass of main-sequence galaxies on redshift and stellar mass:\footnote{We note that the relation in \citet{Tacconi20} also includes a term giving the dependence on the offset from the main-sequence relation. We have dropped this term as our study is restricted to galaxies lying on the main sequence.}

\begin{equation}
\label{eqn:MmolMs}
\log\left[\frac{\Mmol}{\Ms}\right]=A+B\left[\log(1+z)-F\right]^2+D\left[\log\Ms-10.7\right]
\end{equation}
where the coefficients $A=0.06\pm0.2$, $B=-3.33\pm0.2$, $F=0.65\pm0.05$, and $D=-0.41\pm0.03$ \citep[note that the quoted uncertainties are $2\sigma$ errors;][]{Tacconi20}. We will use Equation~\ref{eqn:MmolMs} to determine the $\Mmol-\Ms$ relation at $z\approx0$, $z\approx1.0$, and $z\approx1.3$; these scaling relations are shown as the solid blue lines in Figure~\ref{fig:fig1}[B]. 
		
\subsection{The $\Mat-\Ms$ Relation}
\label{sec:hi}
We determine the $\MHI-\Ms$ relation at $z\approx1.0$ and $z\approx1.3$ from the measurements of the average \hi\ mass of main-sequence galaxies at these redshifts in the GMRT-CAT$z$1 survey \citep{Chowdhury22b}. This survey covered the \hii\ line of 11,419 main-sequence galaxies at $z=0.74-1.45$ with stellar masses $\Ms=10^9-10^{11.5}~\Msun$. \citet{Chowdhury22d} divided the sample of 11,419 galaxies into three stellar-mass subsamples, and  measured the average \hi\ mass of the galaxies in each subsample, and fitted a linear relation to the $\MHI$ and $\Ms$ estimates to determine the $\MHI-\Ms$ relation at $z \approx 1$. They found that the slope of the $\MHI-\Ms$ relation at $z\approx1$ is consistent with that at $z\approx0$. However, the intercept of the relation at $z\approx1$ is higher than that of the local relation by a factor of $\approx3.5$. 

\citet{Chowdhury22d} also determined the $\MHI-\Ms$ relation for blue (NUV$-$r$<4$) galaxies at $z \approx 0$ from $\MHI$ measurements of galaxies covered in the extended GALEX Arecibo Sloan Digital Sky Survey \citep[xGASS;][]{Catinella18}: xGASS is an Arecibo telescope \hii\ survey of a sample of stellar mass-selected galaxies at $z\approx0$, {with stellar masses in the range $10^{9.0} - 10^{11.5} \, \rm M_\odot$}. They obtained the relation $\log[\MHI]=(0.38\pm0.05)\log[{\rm M_{*,10}}] +(9.634\pm0.019)$, where ${\rm M_{*,10}}=\log\left[\Ms/10^{10}\Msun\right]$. Since the slope of the $\MHI - \Ms$ relation for the GMRT-CAT$z$1 galaxies at $z \approx 0.74-1.45$ is the same as that of the relation at $z \approx 0$, we will assume that the slope of the $\MHI - \Ms$ relation is constant, and equal to $0.38$ (i.e. the xGASS slope), over the entire redshift range $z \approx 0 - 1.3$.

Next, \citet{Chowdhury22a} subdivided the DEEP2 galaxies into two redshift bins, at $0.74 < z \leq 1.25$ and $1.25 < z \leq 1.45$, to measure the average \hi\ mass of main-sequence galaxies at average redshifts of $\langle z \rangle \approx 1$ and $\langle z \rangle \approx 1.3$. They obtained average \hi\ masses of $\langle \MHI \rangle = (3.36 \pm 0.64) \times 10^{10}~\Msun$ at $z \approx 1.3$ and $\langle \MHI \rangle = (1.06 \pm 0.19) \times 10^{10}~\Msun$ at $z \approx 1.0$, both for average stellar masses of $\langle \Ms \rangle \approx 10^{10}~\Msun$. We determine the intercepts of the $\MHI-\Ms$ relations at $z\approx1$ and $z\approx1.3$ by requiring that the relations are consistent with the above $\langle \MHI \rangle$ measurements for galaxies with $\langle\Ms\rangle = 10^{10}~\Msun$ at each redshift. This yields the scaling relations 

\begin{equation}
\log[\MHI]=0.38 \log[{\rm M_{*,10}}] +(10.063\pm0.082) \;\;\;\;\; (z\approx1) 
\end{equation}
and 
\begin{equation}
\log[\MHI]=0.38 \log[{\rm M_{*,10}}] + (10.564\pm0.087) \;\;\;\;\; (z\approx1.3)
\end{equation}
 
We note that the above $\MHI-\Ms$ relations yield, at each redshift, the {\it mean} $\MHI$ value for a distribution of galaxies at a given $\Ms$. However, the main-sequence relation of Equation~\ref{eqn:popesso} and the $\MHtwo-\Ms$ relation of Equation~\ref{eqn:MmolMs} yield, respectively, the median SFR and the median $\MHtwo$ at a given stellar mass. For a log-normal distribution of \hi\ masses, which is typically the case \citep[e.g][]{Saintonge22}, the mean \hi\ mass is different from the median \hi\ mass. For consistency with the main-sequence and the $\Mmol-\Ms$ relations, we hence correct our mean $\MHI-\Ms$ relations to obtain the median $\MHI-\Ms$ relations \citep[e.g.][]{Bera22,Chowdhury22d}. Assuming that the \hi\ masses of galaxies at $z\approx1$ and $z\approx1.3$ follow log-normal distributions with a scatter identical to the 0.4~dex scatter measured for the $\MHI-\Ms$ relation at $z\approx0$ \citep{Catinella18}, the median $\MHI-\Ms$ relations lie 0.184~dex lower than the corresponding mean relations. The median $\MHI - \Ms$ scaling relations are then:
\begin{equation}
\log[\MHI]=0.38 \log[{\rm M_{*,10}}] +(9.879\pm0.082) \;\;\;\;\; (z\approx1), 
\end{equation}
and 
\begin{equation}
\log[\MHI]=0.38 \log[{\rm M_{*,10}}] + (10.380\pm0.087) \;\;\;\;\; (z\approx1.3).
\end{equation}

\begin{figure*}
\includegraphics[width=\linewidth]{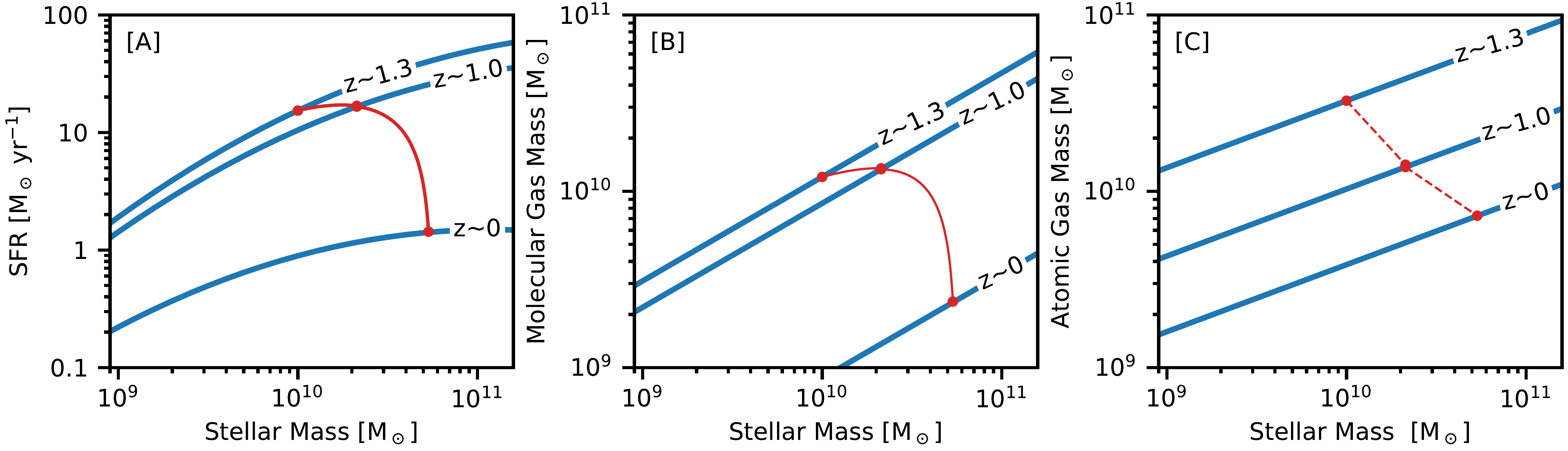}
\caption{The panels show (in blue), as a function of stellar mass,  [A]~the SFR  (Section~\ref{sec:sfr}), [B]~the molecular gas mass (Section~\ref{sec:h2}), and [C]~the atomic gas mass  (Section~\ref{sec:hi}), for main-sequence galaxies at $z\approx0$, $z\approx1.0$ and $z\approx1.3$. The red curve in Panel~[A] shows the evolutionary track of a galaxy with $\Ms=10^{10}~\Msun$ at $z\approx1.3$, obtained using Equation~\ref{eqn:Msbuildup}. The corresponding evolution of the molecular gas mass, obtained using Equations~\ref{eqn:Msbuildup} and \ref{eqn:MmolMs},  is shown in Panel~[B]. Finally, the red points in Panel~[C] shows the atomic gas mass of the galaxy at the three redshifts, obtained using Equation~\ref{eqn:MHIMs}. Note that the $\Mat-\Ms$ relation of Equation~\ref{eqn:MHIMs} is not a continuous function of redshift; the dashed line in the panel is hence only for visual aid, showing a linear interpolation between the points. See main text for discussion. }
\label{fig:fig1}
\end{figure*}

Finally, we also correct the above median relations for the mass contribution of helium, assuming $\Mat=1.36\ \MHI$, to obtain
\begin{equation}
\label{eqn:MHIMs}
\log[\Mat]=0.38 \log[{\rm M_{*,10}}]+A 
\end{equation}
where the values of the coefficient $A$ are $(9.583\pm0.019)$ at $z\approx0$, $(10.012\pm0.082)$ at $z\approx1.0$, and $(10.513\pm0.087)$ at $z\approx1.3$. The solid blue lines in Figure~\ref{fig:fig1}[C] show the median $\Mat-\Ms$ relation at $z\approx0$, $z\approx 1.0$, and $z \approx 1.3$.

{In passing, we note that, over the last two decades, different \hii\ surveys have measured the $\MHI-\Ms$ scaling relation for various galaxy populations at $z\approx0$, each with  different galaxy selection criteria \citep[e.g.][]{Barnes01,Giovanelli05,Huang12,Bothwell13,Denes14,Catinella18,Parkash18}. The scaling relations obtained from ``blind'' \hii\ surveys  \citep[e.g.][]{Barnes01,Giovanelli05} are biased towards \hi-rich galaxies, yielding $\MHI-\Ms$ relations that lie above those obtained from \hii\ surveys of optically-selected galaxies \citep[e.g.][]{Huang12,Catinella18}. The $\MHI-\Ms$ relation at $z\approx1$ from the GMRT-CAT$z1$ survey, as well as the main-sequence relation and the $\MHtwo-\Ms$ relation used in this work, are all obtained from optically-selected galaxy samples. We hence chose to use the scaling relation obtained from the blue galaxies in the optically-selected xGASS sample \citep{Catinella18} as the reference $\MHI-\Ms$ relation at $z\approx0$.}

\section{Results}

We use the formalism of Section~\ref{sec:formalism} with the main-sequence relation (Equation~\ref{eqn:popesso}), the $\Mmol-\Ms$ relation (Equation~\ref{eqn:MmolMs}), and the $\Mat-\Ms$ relation (Equation~\ref{eqn:MHIMs}), to estimate $\Rsfr$, $\Rmol$, and $\Racc$ as a function of stellar mass over the redshift intervals $z\approx1.3-1.0$ and $z\approx1.0-0.0$.
We emphasize that this implicitly assumes that star-forming galaxies evolve along the main sequence \citep[e.g.][]{Renzini09,Peng10,Speagle14,Scoville17}.

\subsection{The Evolution of a Milky-Way-like Galaxy  from $z\approx1.3$ to $z\approx0$}

{We illustrate the formalism of Section~\ref{sec:formalism} for the} case of a massive galaxy with $\Ms=10^{10}~\Msun$ at $z\approx1.3$. We use Equation~\ref{eqn:Msbuildup} and the main-sequence relation of Equation~\ref{eqn:popesso} to evolve the galaxy; the evolutionary track of the galaxy in the SFR$-\Ms$ plane is shown in the red curve in Figure~\ref{fig:fig1}[A]. We find that the stellar mass of the galaxy increases to  $\approx2.1\times10^{10}~\Msun$ at $z\approx1$. Using Equation~\ref{eqn:rsfr}, we find that the time-averaged SFR of the galaxy between $z=1.3$ and $z=1.0$ is $\Rsfr\approx16~\Msun$~yr$^{-1}$. We use the $\Mmol-\Ms$ relation of Equation~\ref{eqn:MmolMs} to find that the molecular gas mass of the galaxy increases from $\Mmol\approx1.2\times10^{10}~\Msun$ at $z=1.3$ to $\Mmol\approx1.3\times10^{10}~\Msun$ at $z=1.0$. Using Equation~\ref{eqn:rhtwo}, this yields a time-averaged molecular gas formation rate of $\Rmol\approx17~\Msun {\rm yr^{-1}}$, very similar to the average SFR, $\Rsfr$, over the same interval. Finally, we use the $\Mat-\Ms$ relation of Equation~\ref{eqn:MHIMs} to find that the atomic gas mass of the galaxy declines from $\Mat\approx3.3\times10^{10}~\Msun$ at $z=1.3$ to $\Mat\approx1.4\times10^{10}~\Msun$ at $z=1.0$. We use Equation~\ref{eqn:racc} to find that the net accretion rate of the galaxy over this interval is $\Racc\approx1~\Msun$~yr$^{-1}$. We thus find that the net accretion rate of a main-sequence galaxy with a stellar mass of $\Ms=10^{10}~\Msun$ at $z\approx1.3$, over the interval $z=1.3$ to $z=1.0$, is far lower than both the average SFR and the molecular gas formation rate over the same period. 

Following the same galaxy down to $z\approx0$, we find that the stellar mass increases to $\Ms\approx5\times10^{10}~\Msun$ (similar to that of the Milky Way), while the molecular gas mass and atomic gas mass both decline, to values of  $\Mmol\approx2\times10^9~\Msun$ and $\Mat\approx7\times10^9~\Msun$, respectively, at $z=0$. We note that that these stellar, atomic, and molecular gas masses are similar to those of the Milky Way \citep[e.g.][]{Kalberla09}. Using Equations~\ref{eqn:rsfr}--\ref{eqn:racc}, we obtain $\Rsfr\approx7~\Msun$~yr$^{-1}$, $\Rmol\approx5~\Msun$~yr$^{-1}$, and $\Racc\approx4~\Msun$~yr$^{-1}$ over the 8-Gyr interval from $z=1.0$ to $z=0$. We thus find that 
both $\Rmol$ and $\Racc$ are lower than $\Rsfr$ during the interval $z=1.0$ to $z=0$; this is unlike the situation in the interval $z=1.3$ to $z=1.0$, where $\Rmol$ was similar to $\Rsfr$.
Further, both $\Rsfr$ and $\Rmol$ are significantly lower in the interval $z = 1.0-0$ than at $z= 1.3 - 1.0$; however, the time-averaged net gas accretion rate $\Racc$ over  $z = 1.0-0$ is higher than that over $z= 1.3 - 1.0$.

\subsection{The Net Gas Accretion  and Molecular Gas Formation Rates as a Function of Stellar Mass}

\begin{figure*}
	\includegraphics[width=\linewidth]{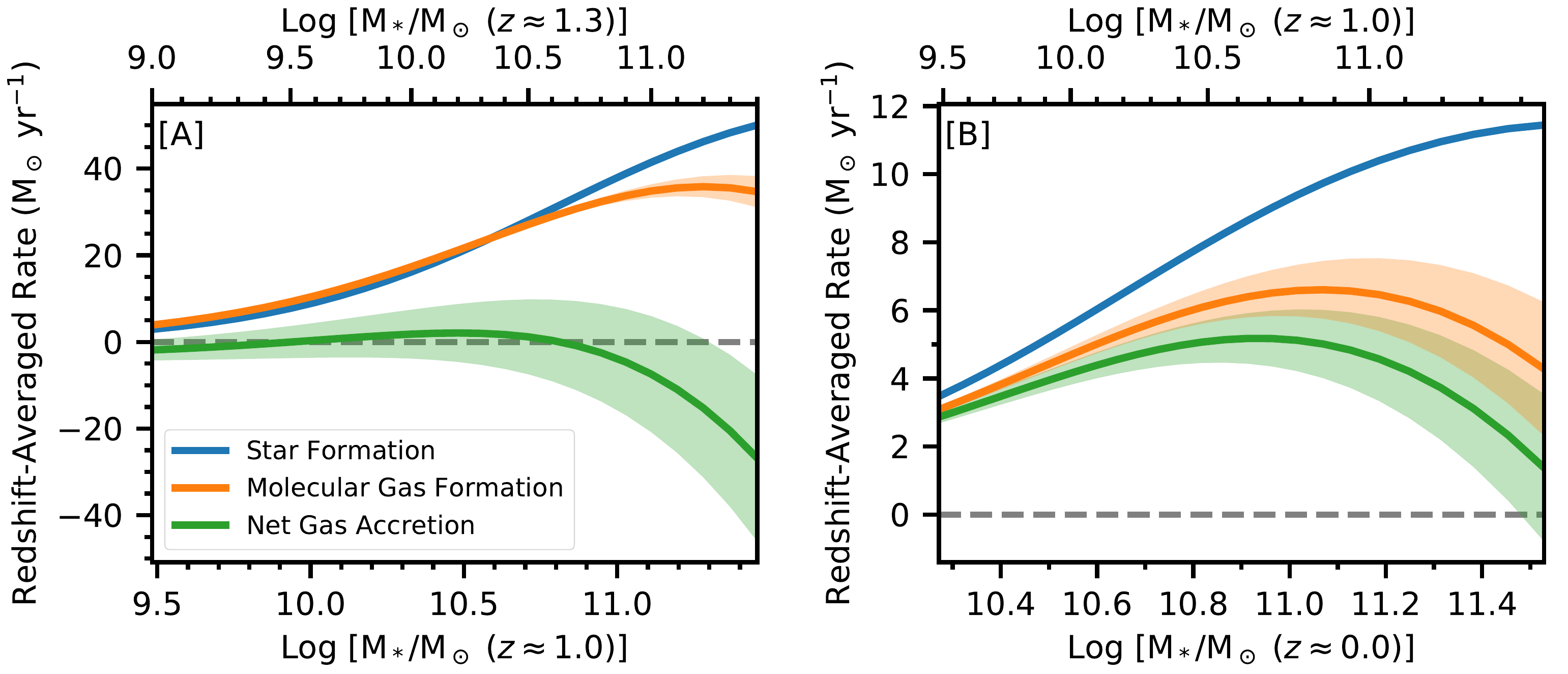}
	\caption{The two panels show the redshift-averaged star-formation rate ($\Rsfr$, in blue), molecular gas formation rate ($\Rmol$, in orange), and net gas accretion rate ($\Racc$, in green) of main-sequence galaxies, as a function of the stellar mass, over [A]~$z=1.3$ to $z=1.0$, and [B]~$z=1.0$ to $z=0$. The shaded regions around each curve show the 68\% confidence intervals. The top x-axis of panel~[A] shows the initial stellar mass of the galaxies at $z=1.3$, while the bottom x-axis shows the final stellar mass, at $z=1.0$. This final stellar mass at $z=1.0$ is plotted as the top x-axis of panel~[B], with the bottom axis of the panel showing the final stellar mass of the galaxies at $z=0$. The evolution of the stellar masses was obtained using Equation~\ref{eqn:Msbuildup}. Panel~[A] shows that the molecular gas formation rate and star-formation rates are similar for all but the most massive galaxies over the interval $z=1.3$ to $z=1.0$, but the net gas accretion rate over the same period is much lower than the other two rates. Panel~[B] shows that both the average molecular gas formation rate and the average net gas accretion rate are lower than the average SFR for the lower-redshift interval, $z=1.0$ to $z=0$, for galaxies of all stellar masses. See main text for discussion.   }
	\label{fig:fig2}
\end{figure*}

{We apply a similar analysis} to main-sequence galaxies with initial stellar masses (at $z =1.3$) in the range $\Ms\approx10^{9}-10^{11.4}~\Msun$, similar to the range of stellar masses over which the GMRT-CAT$z1$ survey obtained the $\MHI-\Ms$ relation \citep{Chowdhury22d}.  Figure~\ref{fig:fig2}[A] shows, as a function of stellar mass at $z=1.0$, our estimates of $\Rsfr$, $\Rmol$, and $\Racc$ over the interval $z=1.3$ to $z=1.0$, while Figure~\ref{fig:fig2}[B] shows the same quantities for the lower redshift interval $z=1.0$ to $z=0$, again plotted against the final stellar mass, at $z = 0$. We note that the confidence intervals shown in Figure~\ref{fig:fig2} take into account the uncertainties in the $\Mat-\Ms$ relation and the $\Mmol-\Ms$ relation, but not that in the main-sequence relation; the uncertainties in our estimates of $\Rmol$ and $\Racc$ are dominated by those in the former two scaling relations.

Figure~\ref{fig:fig2}[A] shows that, except for the most massive galaxies, the time-averaged molecular gas formation rate is similar to the net average SFR over the higher-redshift interval $z=1.3$ to $z=1.0$. Both $\Rsfr$ and $\Rmol$ are $\gtrsim10~\Msun~\textrm{yr}^{-1}$ for galaxies with $\Ms\gtrsim10^{10}~\Msun$ at $z\approx1.3$.  Remarkably, Figure~\ref{fig:fig2}[A] also shows that $\Racc$ for main-sequence galaxies {\it over the entire stellar mass range} is far lower than both $\Rsfr$ and $\Rmol$. Indeed, our estimate of $\Racc$ is consistent with no net accretion of neutral gas onto the disks of star-forming galaxies over the redshift range $z\approx1.3$ to $z\approx1$. This essentially implies that the \hi\ that is consumed in molecular gas formation (and later, star-formation) is not replenished in main-sequence galaxies towards the end of the epoch of galaxy assembly.

For the most massive galaxies, with $\Ms\gtrsim10^{10.5}~\Msun$ at $z\approx1.3$, Fig.~\ref{fig:fig2}[A] indicates that the net gas accretion rate is negative, i.e. that outflows dominate infall in such galaxies over the redshift range $z\approx1.3$ to $z\approx1.0$. Evidence has indeed been found for stronger outflows in higher stellar-mass galaxies at these redshifts: for example, \citet{Weiner09} find that the outflow velocity in DEEP2 galaxies at $z \approx 1.4$ scales $\propto {\rm SFR}^{0.3}$, implying that the outflowing material may escape from high-mass galaxies. However, we emphasize that the result that outflows dominate infall in high-mass galaxies remains tentative, due to the uncertainties in the different scaling relations.

In the lower-redshift interval ($z=1$ to $z=0$), Figure~\ref{fig:fig2}[B] shows that $\Rsfr\approx4-12~\Msun$~yr$^{-1}$, typically far lower (except for the lowest stellar-mass galaxies) than $\Rsfr$ for the same galaxy in the higher redshift interval, $z=1.3$ to $z=0$. This is a direct consequence of the redshift evolution of the main-sequence relation \citep[e.g.][]{Whitaker14,Popesso22}. Further, it can be seen from the figure that $\Rmol$ is lower than $\Rsfr$
for galaxies over the entire stellar-mass range. This is different from the behavior in the higher-redshift interval, where Figure~\ref{fig:fig2}[A] shows that $\Rmol \approx \Rsfr$ for all, except the most massive, galaxies.  Finally, we find that the net average gas accretion rate of main-sequence galaxies over the lower redshift interval is comparable to $\Rmol$ for galaxies with  $\Ms \gtrsim 10^{10.4} \ \Msun$ today; however, $\Racc$ continues to be lower than the average SFR, especially for galaxies today with stellar masses similar to or larger than that of the Milky Way.

\subsection{The Build-up of Baryons in Main-Sequence Galaxies over the past $\approx9$~Gyr}
\begin{figure}
    \centering
    \includegraphics[width=\linewidth]{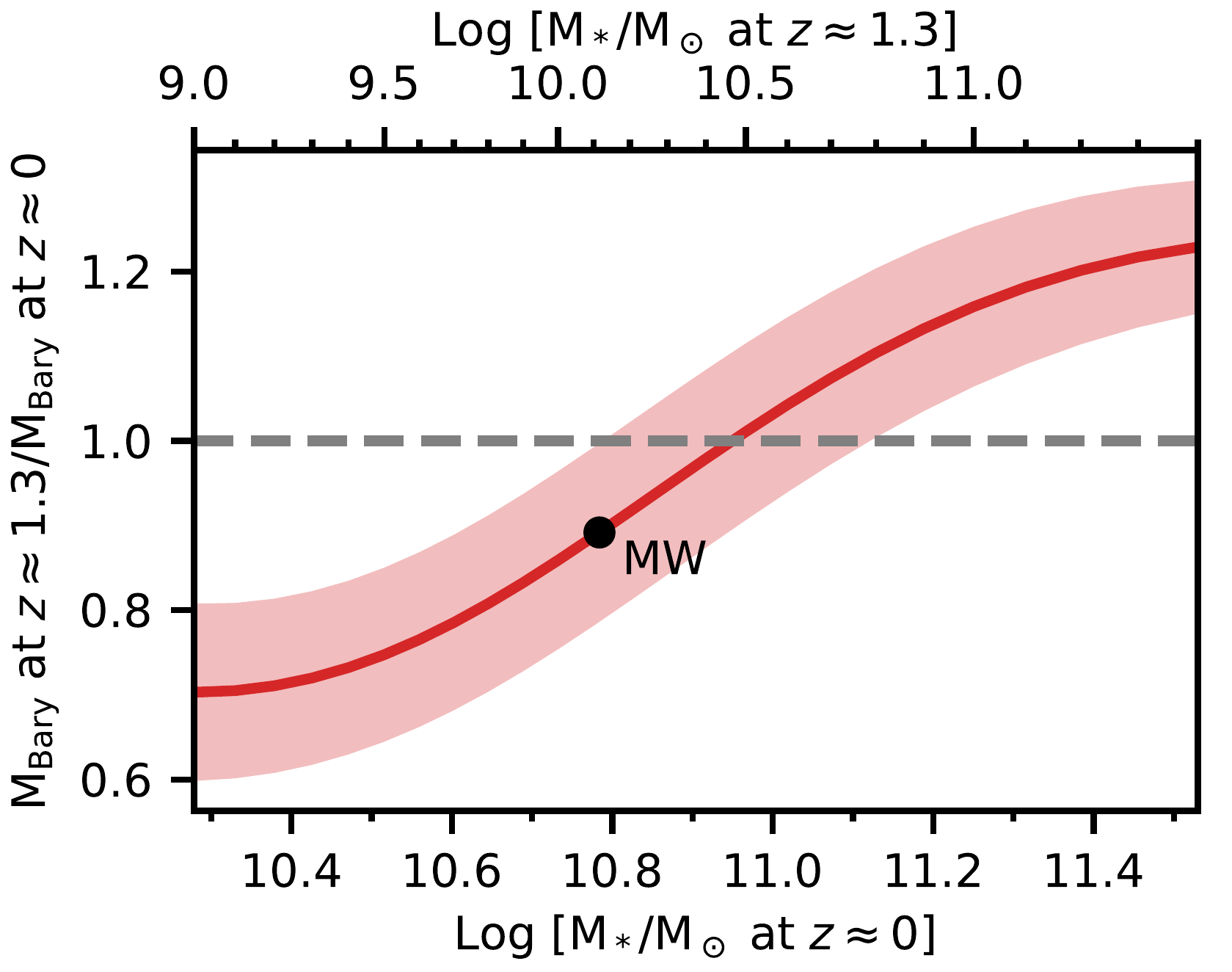}
    \caption{The ratio of the baryonic mass of main-sequence galaxies at $z\approx 1.3$ to that at $z=0$, as a function of their stellar mass at $z\approx0$. The black dot indicates the position on the relation of a galaxy with the stellar mass of the Milky Way today. The dashed horizontal line is at a ratio of 1, indicating no change in the baryonic mass between $z=1.3$ and $z=0$. It is clear that the highest-mass ($\Ms \gtrsim 10^{10.5}~\Msun$ at $z\approx1.3$) galaxies had the bulk of their baryonic mass in place by $z\approx 1.3$; the baryonic mass of these galaxies has declined over the past $\approx 9$~Gyr. Conversely, the baryonic mass of galaxies with stellar mass $\Ms \lesssim 10^{10}~\Msun$ at $z\approx1.3$ has increased since $z \approx 1.3$, by approximately $30\%$ for galaxies with $\Ms \approx 10^{9}~\Msun$ at $z\approx1.3$. }
    \label{fig:fig3}
\end{figure}

{Our formalism can be used to track} the build-up of baryons in galaxies\footnote{We emphasize that this does {\it not} include the baryons in the CGM.} from $z \approx 1.3$ to the present epoch. For this, we combine the estimates of the stellar mass, molecular gas mass, and atomic gas mass to estimate the baryonic mass of main-sequence galaxies at $z\approx1.3$, $z \approx 1$, and $z\approx0$. { Again, we emphasize that we assume that the ionized gas mass within galaxies can be neglected relative to the stellar and neutral gas masses, and that we are not considering the CGM here.} Figure~\ref{fig:fig3} shows, as a function of the stellar mass of galaxies today, the fraction of the baryonic mass in individual galaxies today that was already within the galaxy by $z\approx1.3$. It can be seen that
most of the baryonic mass in star-forming galaxies with $\Ms\gtrsim10^{9}~\Msun$ at $z\approx1.3$ was already in place by $z\approx1.3$. Indeed, Figure~\ref{fig:fig3} indicates that the baryonic mass in the disk of the most massive galaxies, with $\Ms\gtrsim10^{10.4}~\Msun$ at $z\approx1.3$,  has actually \emph{declined over the past $\approx9$~Gyr}, by $\approx 20$\% in galaxies with $\Ms\gtrsim10^{11}~\Msun$ at $z\approx1.3$. The bulk of the evolution in massive galaxies over the last $\approx 9$~Gyr thus appears to involve the conversion of the baryonic material from atomic gas to molecular gas, and thence to stars, rather than acquisition of fresh gas from the CGM. Conversely, gas accretion from the CGM continues to play an important role in the evolution of low stellar-mass galaxies, with $\Ms\lesssim10^{10}~\Msun$ at $z\approx1.3$. Such galaxies have acquired significant amounts of  baryonic material from the CGM over the past $\approx 9$~Gyr, with the largest fractional increase in the baryonic mass ($\approx 30$\%) taking place in the lowest stellar-mass galaxies, 
with $\Ms\approx10^{9}~\Msun$ at $z\approx1.3$

\section{Discussion}

Our results show that the net average gas accretion rate in main-sequence galaxies is lower than the average SFR in both redshift intervals, $z=1.3$ to $z=1.0$, and $z=1.0$ to $z=0$. This implies that the gas accretion rate does not keep pace with the SFR in main-sequence galaxies at $z\lesssim 1.3$: the star-formation activity in such galaxies is driven by the \hi\ that is present in their disks from earlier times. Our results thus suggest that the net gas accretion was significantly more efficient at high redshifts, $z \gtrsim 1.3$, causing the build-up of \hi\ in the disks of main-sequence galaxies. It is plausible that this efficient accretion is via the cold mode, with gas flowing along filaments onto the disks of galaxies \citep[e.g.][]{Keres05,Dekel09}.  In this scenario, the large \hi\ reservoirs resulted in a very efficient conversion of \hi\ to \htwo\ at $z\approx1-3$, consistent with our estimate of a high $\Rmol$, comparable to $\Rsfr$, over $z\approx1.3-1.0$. This efficient \hi-to-\htwo\ conversion, in turn, triggers the high star-formation activity in galaxies at these redshifts,
resulting in the observed era of high star-formation activity over $z \approx 1-3$, the epoch of galaxy assembly \citep{Madau14}

At later times, $z \lesssim 1.3$, a combination of physical processes are likely to have led to the observed low net gas accretion rate of Figure~\ref{fig:fig2}[B]. These include a transition to inefficient hot-mode accretion in  massive galaxies, due to the increase in the masses of their dark matter halos \citep[e.g.][]{Dekel06}, an increased efficiency of stellar- and AGN-driven outflows in massive galaxies that remove gas from galaxies \citep[e.g.][]{Weiner09,Veilleux20}, interactions between neighbouring galaxies that disrupt the smooth cold-mode filamentary flows, etc. The low net gas accretion rate at $z \lesssim 1.3$, combined with the very high SFR and high molecular gas formation rate, implies that the \hi\ reservoir is consumed on a very short timescale of $\lesssim 1$~Gyr \citep{Chowdhury22b}. This results in a decline in the average \hi\ content of galaxies by $z \approx 1$ \citep{Chowdhury22a}, and  in a lower average molecular gas formation rate at later times, as seen in Figure~\ref{fig:fig2}[B] at $z < 1$. The resulting decline in the amount of molecular gas available for star formation causes a decline in the star-formation activity in individual galaxies, especially in the more massive systems that dominate the SFR density of the Universe at $z \gtrsim 1$. {\it The low net gas accretion rate at $z \lesssim 1.3$ is thus the primary cause of the observed decline in the SFR density of the Universe at $z \lesssim 1$.} 
We emphasize that the above scenario is consistent with the measurements of the large \hi\ and \htwo\ reservoirs of main-sequence galaxies at $z\gtrsim1$, and the decline in the neutral-gas content of these galaxies at later times \citep{Chowdhury22a,Chowdhury22b,Chowdhury22c,Tacconi20}.

{ Our estimates of $\Rsfr$, $\Rmol$, and $\Racc$ over $z\approx0-1$ are the time-averaged rates over a relatively long, $\approx8$~Gyr, interval. This implies that there could be significant evolution in each of these quantities between $z\approx1$ and $z\approx0$ that is not captured by the averages in Figure~\ref{fig:fig2}[B]. Indeed, \citet{Bera23b} apply a similar formalism to find that galaxies over the last $\approx4$~Gyr, from $z\approx0.35$ to $z\approx0$, have $\Racc \approx \Rsfr$, while $\Rmol$ is lower than the other two\footnote{We note, for ease of comparison, that Figure~4 of \citet{Bera23b} shows the rate of change of the molecular gas content of galaxies, i.e. $(\Mmolf-\Mmoli)/\Delta t \equiv (\Rmol - \Rsfr)$, rather than $\Rmol$ as defined in Equation~\ref{eqn:molgas}; similar conclusions are reached using either metric.}. They hence conclude that the decline in the star-formation activity over the past $\approx 4$~Gyr is primarily due to inefficient conversion of \hi\ to \htwo\ \citep{Bera23b}. Combining the two results, it then appears that for the first few Gyrs after $z\approx1$, the star-formation activity continues to be limited by the accretion of \hi\ onto galaxies, but at later times, even though the relatively low SFRs of galaxies are now comparable to their gas accretion rates, the star-formation activity continues to decline, but now primarily due to the low molecular gas formation rate. However, we caution that. the ranges of stellar mass explored in the study of \citet{Bera23b} and this work are very different, with the typical stellar masses here being far higher than those in \citet{Bera23b}.  In addition, as pointed out by \citet{Bera23b}, the small cosmic volume covered in their study implies that their $\MHI - \Ms$ scaling relation may be affected by cosmic variance.}
\hfill\\

\section{Summary}

{In this Letter,} we have combined our measurement of the $\MHI-\Ms$ scaling relation in main-sequence galaxies at $z \approx 0.74-1.45$ from the GMRT-CAT$z$1 survey, with estimates of the main-sequence relation and the $\Mmol-\Ms$ relation, and the assumption of the continuity of main-sequence evolution, to determine the net average gas accretion rate and the molecular gas formation rate in main-sequence galaxies over two interesting intervals, $z=1.3$ to $z=1.0$, at the end of the epoch of galaxy assembly, and $z=1.0$ to $z=0$, the period when the SFR density of the Universe declines by an order of magnitude. We find that the average molecular gas formation rate is comparable to the average SFR in the former interval, but that the net gas accretion rate is significantly lower than both $\Rmol$ and $\Rsfr$. The low net gas accretion rate results in a depletion of the \hi\ reservoir in massive galaxies by $z \approx 1$, and causes the decline in the SFR density of the Universe at lower redshifts. In the lower-redshift interval, both $\Racc$ and $\Rmol$ are lower than $\Rsfr$ for all stellar masses; this results in the consumption of both the \hi\ and \htwo\ reservoirs, and a continuing decline in the SFR density of the Universe down to the present epoch. We also find that most of the baryonic contents of massive main-sequence galaxies, with $\Ms \gtrsim 10^{10.5}~\Msun$ at $z\approx1.3$, was already in place $\approx 9$~Gyr ago, while low-mass galaxies, with $\Ms \lesssim 10^{10}~\Msun$ at $z\approx1.3$, have built up approximately $30\%$ of their baryonic content over this period.

\begin{acknowledgments}
We thank an anonymous referee for a report that improved the clarity of this paper. We also thank the staff of the GMRT who have made these observations possible. The GMRT is run by the National Centre for Radio Astrophysics of the Tata Institute of Fundamental Research. AC and NK thank Apurba Bera for many discussions about \hi\ scaling relations and gas accretion that have contributed to this work. NK acknowledges support from the Department of Science and Technology via a Swarnajayanti Fellowship (DST/SJF/PSA-01/2012-13). AC, NK, $\&$ JNC also acknowledge the Department of Atomic Energy for funding support, under project 12-R\&D-TFR-5.02-0700. 
	\end{acknowledgments}
    \software{      astropy \citep{astropy:2013}}
      
    \bibliography{bibliography.bib}

\bibliographystyle{aasjournal}

\end{document}